\documentclass[fleqn,10pt]{article}
\usepackage{amsmath}
\usepackage{amssymb}
\usepackage{amsthm}
\usepackage{array}
\usepackage{comment}
\usepackage{dirtytalk}
\usepackage{enumitem}
\usepackage{graphics}
\usepackage{geometry}
\usepackage{mathrsfs}
\usepackage{multirow}
\usepackage{tikz}
\usepackage{tkz-graph}
\usepackage{xcolor}
\usetikzlibrary{arrows,calc,shapes}
\makeatletter

\newcommand*{\itemequation}[3][]{%
	\item
	\begingroup
	\refstepcounter{equation}%
	\ifx\\#1\\%
	\else  
	\label{#1}%
	\fi
	\sbox0{#2}%
	\sbox2{$\displaystyle#3\m@th$}%
	\sbox4{\@eqnnum}%
	\dimen@=.5\dimexpr\linewidth-\wd2\relax
	\ifcase
	\ifdim\wd0>\dimen@
	\z@
	\else
	\ifdim\wd4>\dimen@
	\z@
	\else 
	\@ne
	\fi 
	\fi
	\@latex@warning{Equation is too large}%
	\fi
	\noindent   
	\rlap{\copy0}%
	\rlap{\hbox to \linewidth{\hfill\copy2\hfill}}%
	\hbox to \linewidth{\hfill\copy4}%
	\hspace{0pt}
	\endgroup
	\ignorespaces 
}
\newcolumntype{P}[1]{>{\centering\arraybackslash}p{#1}}

\newtheorem{definition}{Definition}
\newtheorem{example}{Example}

\newcommand{\specialcell}[2][c]{\begin{tabular}[#1]{@{}c@{}}#2\end{tabular}}

\tikzstyle{decision} = [diamond, draw, fill=blue!20, text width=4.5em, text badly centered, node distance=3cm, inner sep=0pt]
\tikzstyle{block} = [rectangle, draw, fill=blue!20, text width=5em, text centered, rounded corners, minimum height=4em]
\tikzstyle{line} = [draw, -latex']
\tikzstyle{cloud} = [draw, ellipse,fill=red!20, node distance=3cm, minimum height=2em]

\title{Arguing Ecosystem Values with Paraconsistent Logics}

\author{Juan Afanador \\ University of Edinburgh}
\providecommand{\keywords}[1]{\textbf{\textit{Keywords}:} #1}

\begin{document}
	
	\flushbottom
	\maketitle
	\thispagestyle{empty}
	
	\begin{abstract}
		The valuation of ecosystem services prompts dialogical settings where non-trivially inconsistent arguments are often invoked. Here, I propose an approach to the valuation of ecosystem services circumscribed to a logic-based argumentation framework that caters for valid inconsistencies. This framework accounts for preference formation processes underpinned by a paraconsistent model of logical entailment. The value of an ecosystem service is produced in the form of an ordering over competing land-use practices, as per the arguments surviving semantical probing.
	\end{abstract}

	\keywords{valuation of ecosystem services, abstract argumentation, paraconsistent logics, Dialetheism.}
	
	\section{Introduction}
	
	The valuation of ecosystem services deals with distinct and contradictory views on value, a multiplicity that typically reflects opposing or conflicting land-use practices. At times, it is the case that the same people who acknowledge the importance of the benefits derived from a particular ecosystem function, and who are also aware of the connection between the two, undertake actions that contravene the precedence of such knowledge. Furthermore, these decisions are not made in the abstract, but mediated by various forms of sociality involving dialogue ---even if only implicitly.
	
	This document presents an approach to the valuation of ecosystem services that incorporates the contradictions and inconsistencies inherent in valuation ---an approach I term \textit{Argumentative Valuation} (AV). AV relies on an argumentation framework structured around Dialetheism. AV's argumentation framework enables the dialogical settings where values substantiate, while its dialetheic substrate allows for true (truthful) inconsistencies in the values' antecedents. 
	
	The presentation will be restricted to the more technical aspects of my proposal and their interplay in a unified analytical framework. Other considerations related to the discussion of a situated notion of value, and the adequacy of AV therewith are, although following an admittedly artificial separation, treated  elsewhere. Thus, without further preamble, AV's abstract argumentation framework will be delineated in Section 2. Section 3 introduces AV by means of a (classical logic-based) example, Section 4 describes AV's logical foundation, and Section 5 summarises our main conclusions and themes for future work. 
	
	\section{Computational Argumentation}
	\label{sec:argumentation}
	
	Computational argumentation is defined as the formal modelling of a chain of arguments interacting in a dialogical setting \cite{Walton2009}. The arguments that uphold or contradict a particular position are compared against each other and probed for flaws. Since every argument is composed of a premise that implicates a conclusion, this process involves identifying the assumptions and associated claims that can be defeated with the use of their own set of supporting premises \cite{besnard2009argumentation}. 
	
	Dung's model of argumentation is the current paradigm of computational argumentation in Artificial Intelligence \cite{Walton2009}. Also referred to as an \textit{abstract argumentation framework} (AF) it is defined as follows:
	
	\begin{definition}[\textbf{Abstract Argumentation Framework} \cite{dung1995acceptability}\label{def:af}] An abstract argumentation framework is a pair $\mathcal{AF}=\langle \mathcal{A},\mathbf{R}\rangle$. $\mathcal{A}$ is a set of arguments and $R\subseteq \mathcal{A}\times\mathcal{A}$ is a binary relation of \textit{attack}. An argument $A\in\mathcal{A}$ \textit{defeats} an argument $B\in\mathcal{A}$ iff $(A,B)\in\mathbf{R}$. A set $\mathcal{S}$ of arguments is said to defeat an argument $A$ iff some argument in $\mathcal{S}$ defeats $\mathcal{A}$.
	\end{definition}
	
	The semantics of an AF is posited in terms of the framework's relation of attack $\mathbf{R}$. $\mathbf{R}$ effectively delineates the criteria to determine the acceptability of a particular argument with respect to distinct arguments. These criteria form the basis for the critical questioning of arguments; a process that requires establishing which subsets of arguments are conflict-free (Definition \ref{def:conflict_free}), in order to derive the (preferred/stable/grounded) extensions of the corresponding AF (Definition \ref{def:accept_semantic}), representing the ultimate collection of tenable or defensible arguments within the framework.
	
	\begin{definition}[\textbf{Conflict-free Sets of Arguments} \cite{prakken2011overview}\label{def:conflict_free}] Let $\mathcal{B}\subseteq\mathcal{A}$ be a subset of arguments of $\mathcal{AF}=\langle \mathcal{A},\mathbf{R}\rangle$. $\mathcal{B}$ is \textit{conflict-free} iff there exist no $A_i, A_j\in\mathcal{B}$ such that $(A_i,A_j)\in\mathbf{R}$. $\mathcal{B}$ \textit{defends} an argument $A_i$ iff for each $A_{j\neq i}\in\mathcal{A}$, there exists $A_k\in\mathcal{B}$ such that $(A_k,A_j)\in\mathbf{R}$, whenever $(A_j,A_i)\in\mathbf{R}$.
	\end{definition}
	
	\begin{definition}[\textbf{Acceptability Semantics} \cite{prakken2011overview}\label{def:accept_semantic}] Let $\mathcal{B}\subseteq\mathcal{A}$ be a subset of arguments of $\mathcal{AF}=\langle \mathcal{A},\mathbf{R}\rangle$. 
		\begin{itemize}[leftmargin=.25in]
			\item $\mathcal{B}$ is an \textit{admissible} set iff $\mathcal{B}$ is conflict-free, and $\mathcal{B}$ is defends all of its arguments.
			\item $\mathcal{B}$ is a \textit{preferred extension} iff $\mathcal{B}$ is maximal admissible set w.r.t set-inclusion.
			\item $\mathcal{B}$ is a \textit{stable extension} iff $\mathcal{B}$ is conflict-free, and $\mathcal{B}$ attacks all the arguments in $\mathcal{A}\setminus\mathcal{B}$.
			\item $\mathcal{B}$ is a \textit{complete extension} iff $\mathcal{B}$ is admissible and contains all arguments it defends.
			\item $\mathcal{B}$ is a \textit{grounded extension} iff $\mathcal{B}$ is \textit{conflict-free}, and $\mathcal{B}$ is the least complete extension w.r.t set-inclusion.
		\end{itemize}
	\end{definition}
	
	The notion of extension can be refined by labelling the arguments in a generic $\mathcal{AF}=\langle \mathcal{A},\mathbf{R}\rangle$, depending on whether they are defeated in $\mathbf{R}$ or not. An argument's status or \textit{label} is \textit{in} iff all arguments defeating the former argument are \textit{out}; conversely, an argument is labelled \textit{out} iff it is defeated by an argument labelled as \textit{out}. Thus, if we define $Undecided(\mathcal{B}):=\mathcal{A}\setminus(In\cup Out)$, where $In:=\{A|A \text{ is \textit{in} } \forall A\in\mathcal{B}\}$ and $Out:=\{A|A \text{ is \textit{out} } \forall A\in\mathcal{B}\}$ for some set $\mathcal{B}\subseteq\mathcal{A}$ of candidate arguments, the extensions of $\mathcal{AF}$ acquire a more terse representation:
	
	\sloppy\begin{definition}[\textbf{Acceptability Semantics With Labels} \cite{cerutti2014}\label{def:label_semantic}] Let $\mathcal{B}\subseteq\mathcal{A}$ be a subset of arguments of $\mathcal{AF}=\langle \mathcal{A},\mathbf{R}\rangle$. 
		\begin{description}[labelindent=.25in]
			\item[Stable Extension:] $\mathcal{B}$ is a stable extension iff $Undecided(\mathcal{B})=\emptyset$.
			\item[Preferred Extension:] $\mathcal{B}$ is a preferred extension iff $Undecided(\mathcal{B})=max_{\mathcal{C}\in\mathcal{P}(In\cup Out)}|\mathcal{C}|$.
			\item[Grounded Extension:] $\mathcal{B}$ is a grounded extension iff $Undecided(\mathcal{B})=min_{\mathcal{C}\in\mathcal{P}(In\cup Out)}|\mathcal{C}|$
		\end{description}{}
	\end{definition}
	
	With Definition \ref{def:label_semantic} at hand, we could further qualify an argument's status:
	
	\sloppy\begin{definition}[\textbf{Acceptability Status of an Argument} \cite{cerutti2014}\label{def:accept_state}] For grounded/stable/preferred semantics, an argument $A\in\mathcal{B}\subseteq\mathcal{A}$, is \textit{justified} iff $A\in In$, \textit{overruled} iff $A\in Out$, or \textit{defensible} iff $A\in Undecided(\mathcal{B})$.
	\end{definition}
	
	Example \ref{ex:extensions} illustrates how Definitions \ref{def:af}--\ref{def:accept_state} can be used to derive a set of defensible arguments. 
	\sloppy\begin{example}[\textbf{The Odd Defeat Loop} \cite{prakken2011overview}\label{ex:extensions}] Figure \ref{fig:example_extensions} depicts the (semi-formal) argumentation tree of the abstract framework $\mathcal{AF}_O=\langle\{A,B,C,D\},\{(A,B),(B,A),(C,A),(C,B),(C,D)\}\}$. Its grounded extension is empty, while the two preferred (and stable) extensions correspond to $P_A=\{A,D\}$ and $P_B=\{B,D\}$. That is, in grounded semantics all arguments are defensible, and in preferred and stable semantics $A$ and $B$ are defensible, while $D$ is justified and $C$ is overruled.
		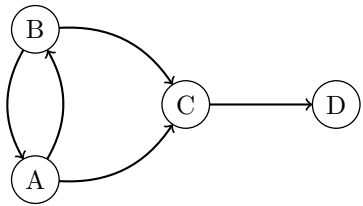
\begin{figure}[h]
			\centering
			\begin{tikzpicture}[x=2cm,y=1cm]
			\SetUpEdge[style={ultra thick}, color=black]
			\Vertex{C}
			\EA(C){D}
			\NOWE(C){B}
			\SOWE(C){A}
			\Edge[style={->, bend right}](A)(C)
			\Edge[style={->, bend right}](A)(B)
			\Edge[style={->, bend right}](B)(A)
			\Edge[style={->, bend left}](B)(C)
			\Edge[style={->}](C)(D)
			\end{tikzpicture}
			\caption{Argumentation Tree for $\mathcal{AF}_O$}
			\label{fig:example_extensions}
		\end{figure}{}
	\end{example}{}
	
	Example \ref{ex:extensions} not only implements the concepts reviewed so far, it also brings out the importance of selecting one or another semantics. In this respect, abstract argumentation frameworks may sometimes seem overly abstract, for however useful Definitions \ref{def:conflict_free}--\ref{def:label_semantic} may be in determining the existence of a choice from a set of practical arguments, they do not provide any insights into the motivations behind said choice or the possibility of predicting future decisions. Value-based argumentation frameworks (VAFs), on the other hand, incorporate this additional information into AFs \cite{bench2009abstract}, while making preferred semantics the natural semantics to derive extensions in situations where preferences are deemed crucial to the critical assessment of an argument.
	
	\begin{definition}[\textbf{Value-based Argumentation Framework} \cite{bench2009abstract}\label{def:vaf}] A value-based abstract argumentation framework is a tuple $\mathcal{VAF}=\langle \mathcal{A},\mathbf{R},V,val,P\rangle$. $\mathcal{A}$ is a set of arguments, $R\subseteq \mathcal{A}\times\mathcal{A}$ is a binary relation of \textit{attack}, $V$ is a non-empty set of \textit{values}, $P$ is the set of total orders on $V$, and $val:\mathcal{A}\mapsto V$. 
	\end{definition}
	
	The AF notions of defeat, acceptability and connected ideas undergo minor changes when applied to VAFs. They have to be adjusted for the axiological nature of the latter, as to produce more informed extensions. This involves complementing the information expressed through $\mathbf{R}$ with the information from $val(\cdot)$, while recognising the distinct preferences contained in the multiple orders ---or \textit{audiences}--- that make up $P$.
	
	\begin{definition}[\textbf{Preferences in VAFs}\label{def:vaf_pref}]
		The preferences of an audience $a\in P$ are represented as a transitive, irreflexive and asymmetric relation $\mathit{Valpref_a}\subseteq V\times V$.
	\end{definition}{}
	
	\begin{definition}[\textbf{Defeat for an Audience} \cite{bench2009abstract}\label{def:def_aud}]
		An argument $A\in\mathcal{A}$ \textit{defeats-for-$a$} an argument $B\in\mathcal{A}$, with respect to an audience $a\in P$, iff $\mathbf{R}(A,B)$ and not $(val(B), val(A))\in \mathit{Valpref_a}$. 
	\end{definition}{}
	
	\begin{definition}[\textbf{Acceptable to an Audience} \cite{bench2009abstract}\label{def:accept_aud}]
		An argument $A\in\mathcal{A}$ is \textit{acceptable-to-$a$}, with respect to an audience $a\in P$ and a set of arguments $\mathcal{B}\subset\mathcal{A}$, if there exists an argument in $\mathcal{B}$ which defeats-for-$a$ any argument in $\mathcal{A}$ that defeats-for-$a$ argument $A$. 
	\end{definition}{}
	
	\begin{definition}[\textbf{Conflict-free for an Audience}\label{def:conflict_aud}]
		A set of arguments $\mathcal{B}\subset\mathcal{A}$ is \textit{conflict-free-for-$a$}, with respect to an audience $a\in P$,
		iff there exist no $A_i, A_j\in\mathcal{B}$ such that $(A_i,A_j)\in\mathbf{R}$ or $(val(A_i), val(A_j)\in\mathit{Valpref}_a$. 
	\end{definition}{}
	
	\begin{definition}[\textbf{Admissible for an Audience} \cite{bench2009abstract}\label{def:adm_aud}]
		An argument $A\in\mathcal{A}$ is \textit{admissible-for-$a$}, with respect to an audience $a\in P$, if every argument in a conflict-free-for-$a$ set of arguments is also acceptable-to-$a$ within the same set.
	\end{definition}{}
	
	The definitions of justified, overruled and defensible arguments are similarly adjusted to the value-based framework. With these and Definitions \ref{def:vaf_pref}--\ref{def:adm_aud} in mind, we can obtain the set of preferred arguments in a given VAF by simply looking at the maximal admissible-for-$a$ subsets, within the corresponding set of arguments. Equivalently:
	
	\begin{definition}[\textbf{Preferred Extension for an Audience} \cite{bench2009abstract}\label{def:ext_aud}]
		A set of arguments $\mathcal{B}\subset\mathcal{A}$ such that $\mathcal{B}=max_{\mathcal{C}\in\mathcal{P}(In\cup Out)}|\mathcal{C}|$, where $In:=\{A|A \text{ is \textit{admissible-for-$a$} } \forall A\in\mathcal{B}\}$ and $Out:=A\setminus In$, is a \textit{preferred-extension-for-$a$}. 
	\end{definition}{}
	
	Although the operation of a fully formalised VAF is deferred to Example \ref{ex:main}, it is already apparent that the VAF approach is central to AV because of its readiness to incorporate values and preferences. Arguments are no longer accepted on the basis of an exogenous heuristic, but contrasted in accordance with the desiderata of their originating agents. Notwithstanding, the VAF approach remains too intuitionistic to integrate the qualitative and quantitative aspects of valuation \cite{bench2007audiences}, and ill-equipped to handle its inconsistencies \cite{dunne2004complexity}, for these challenges occur at the more fundamental level of argument formation. Hence my insistence on the explicit formalisation of the deductive mechanism underpinning VAFs.
	
	\section{(Classical Logic-Based) Argumentative Valuation}
	\label{sec:example}
	Argumentative Valuation is based on the logical formalisation of dialogical interactions. As originally envisioned, its logical formalism should be non-classical. That is, the logic of Argumentative Valuation should contain non-trivial theses of the form $\{a \land \neg a\}$, where $a$ is a premise belonging to the support of a generic argument $A\in\mathcal{A}$. 
	
	Current research into logics amenable to structural contradictions ---within the computational argumentation domain--- is still incipient \cite{amgoud2010formal, arioua2017logic, prakken2018historical}. For this reason, rather than presenting the actual logic with which Argumentative Valuation operates, this section elucidates what is meant by model of logical entailment in the context of my proposal, and how it is articulated into the abstract argumentation framework by means of an example. The justification of a non-classical model of entailment is given in Section \ref{sec:par_logics}, whereas its integration into a working valuation framework will be investigated in future endeavours. Let us, then, introduce a classical logic-based instance of Argumentative Valuation through Example \ref{ex:main}.
	
	\begin{example}[\textbf{Agriculture vs. Restoration in Paramo Sites}\label{ex:main}]
		Our example represents the predicament between using paramo sites for agricultural purposes and performing restoration activities involving the same areas. Our variables of interest are designated as follows:
		
		\begin{description}
			\item[$a:$] Increased agricultural land-use.
			\item[$h:$] Improved hydrological regulating services.
			\item[$r:$] Greater number of peatland restoration activities.
			\item[$s:$] Increased water supply.
			\item[$y:$] Greater agricultural yield.
			\item[$w:$] Improved living conditions.
		\end{description}
		
		To construct our logic-based VAF we make use of sentential logic with a conventional linguistic structure. The resulting construct is considered a classical logic \cite{buning1999propositional}. This classical logic consists of well-formed formulae built up from sentential variables ---such as $h$ or $y$--- using the sentential connectives $\to$ (implies), $\land$ (and), $\lor$ (or), and $\neg$ (not). Every argument in our VAF is formed by combining these sentential elements.
		
		We start off by defining a knowledge base, which summarises all relevant information regarding our variables:
		
		$$KB_0 := \{a, a\to \neg h,a\to y, h\to w, r, r\to\neg a, r\to\neg y, r\to h, y\}$$
		
		The main or \textit{root} argument will be that of prioritising agricultural land-use, as it leads to a greater yield (and higher income), i.e.,
		
		$$A_1 := \langle \{a, a\to y\}, y\rangle.$$
		
		In contrast, restoration activities could be directly advocated ---agricultural land-use is eschewed in favour of peatland restoration activities to improve water quality, i.e.,
		
		$$A_2 := \langle\{r, r\to \neg a\}, \neg(a\land (a\to y))\rangle;$$
		
		which, in turn, could be refuted by the primacy of the income from agricultural activities over non-monetary considerations, i.e.,
		
		$$A_3 := \langle\{y, y\to \neg r\}, \neg (r\land (r\to \neg y))\rangle.$$
		
		Yet, another argument may affirm that restoration is desirable as it guarantees the regulating services hampered by agricultural activities, i.e.
		
		$$A_4 := \langle\{(r\to h)\to (h\to \neg a)\}, \neg(a\land (a\to y))\rangle.$$
		
		In consequence, the abstract argumentation framework of our example takes the form
		
		$$\mathcal{AF}_{KB_0}=\{\{A_1, A_2, A_3, A_4\}, \{(A_1, A_2), (A_1, A_4), ( A_2, A_3)\}\};$$
		
		and its corresponding argumentation tree appears as in Figure \ref{fig:arg_tree_1}.
		
		\begin{figure}[h]
			\centering
			\begin{tikzpicture}[sibling distance=6cm, every node/.style = {shape=rectangle, rounded corners, draw, align=center}, line width=.3mm]]
			\node {$\langle \{a, a\to y\}, y\rangle$}
			child { node {$\langle\{r, r\to \neg a\}, \neg(a\land (a\to y))\rangle$} edge from parent[<-]
				child { node {$\langle\{y, y\to \neg r\}, \neg (r\land (r\to \neg y))\rangle$} edge from parent[<-]}}
			child { node {$\langle\{(r\to h)\to (h\to \neg a)\}, \neg(a\land (a\to y))\rangle$} edge from parent[<-]};
			\end{tikzpicture}    
			\caption{Argumentation Tree for $\mathcal{AF}_{KB_0}$}
			\label{fig:arg_tree_1}
		\end{figure}
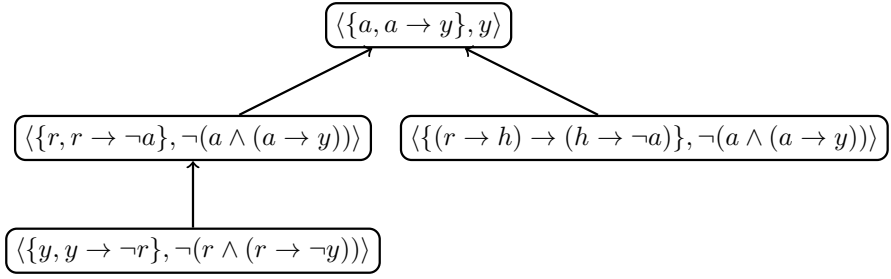{}
		
		To extend our current AF to a value-based framework we recognise $V=\{y, w\}$ as our set of values, in view that $a\to y$ and $(r\to h)\to (h\to w)$. So, there are two possible audiences $P=\{y\succ w, w\succ y\}$ with respect to the binary relation $\succ$ on $V$, which we designate as audience-$y$ associated with order $y\succ w$ and audience-$w$ associated with order $w\succ y$. Their respective variables are tagged with the subscript $(\cdot)_i$ for $i\in\{y,w\}$, whenever the latter's omission may seem ambiguous. Finally, note that $val=\{A_1\mapsto y, A_2\mapsto w, A_3\mapsto y, A_4\mapsto w\}$ and our VAF can be expressed as follows\footnote{The symbol ``$\mapsto$" is used  to represent the mapping from arguments to values, as per Definition \ref{def:vaf}. It is distinct from the implication ``$\to$".}
		
		{\small\begin{align}
			\begin{split}
			\mathcal{VAF}_{KB_0}&=\{\{A_1, A_2, A_3, A_4\}, \{(A_1, A_2), (A_1, A_4), (A_2, A_3)\}, \{y, w\},\\
			& \{A_1\to y, A_2\to w, A_3\to y, A_4\to w\}, \{y\succ w, w\succ y\}\}
			\end{split}
			\end{align}}
		
		The Hasse diagrams in Figure \ref{fig:hasse_1} offer an alternative representation of the preferred extensions of $\mathcal{VAF}_{KB_0}$. It indicates that the preferences contained in our VAF are uninformative, as we end up with two independent sets of preferred arguments for each audience. Put another way the values associated with these arguments are trivial: audience-$y$ favours agricultural activities over restoration as opposed to audience-$w$, precluding the quest for a common ground. However, if the knowledge base is extended to include premises stating that ``increased water supply generates greater agricultural yields", and that ``improved regulating services uphold an increased water supply", i.e., $KB_1\equiv KB_0\cup\{s\to y, h\to s\}$, then a new preferred extension arises.
		
		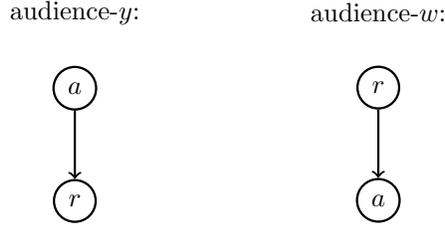
\begin{figure}
			\centering
			\begin{tikzpicture}[sibling distance=6cm, line width=.3mm]]
			\node [circle, draw] (o) {$a$}
			child {node [circle, draw] {$r$} edge from parent[->]};
			\node[above of= o,fill=white,inner sep=.5pt, line width=2pt] {audience-$y$:};
			\end{tikzpicture}
			\hspace{2cm}
			\begin{tikzpicture}[sibling distance=6cm, line width=.3mm]]
			\node [circle, draw] (o) {$r$}
			child {node [circle, draw] {$a$} edge from parent[->]};
			\node[above of= o,fill=white,inner sep=.5pt, line width=2pt] {audience-$w$:};
			\end{tikzpicture}
			\caption{Hasse Diagram for $\mathcal{VAF}_{KB_0}$}
			\label{fig:hasse_1}
		\end{figure}{}
		
		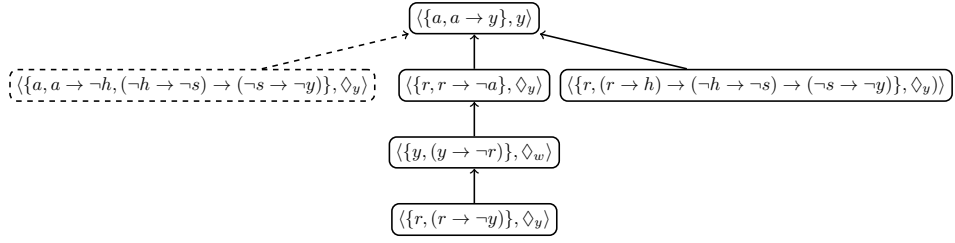
\begin{figure}
			\centering
			\scalebox{.7}{
				\begin{tikzpicture}[sibling distance=6.3cm, every node/.style = {shape=rectangle, rounded corners, draw, align=center}, line width=.3mm]
				\begin{scope}[scale=.85]
				\node {$\langle \{a, a\to y\}, y\rangle$}
				child { node[dashed] {$\langle\{a, a\to \neg h, (\neg h\to\neg s)\to (\neg s\to\neg y)\}, \lozenge_y\rangle$} edge from parent[<-, dashed]}
				child { node {$\langle\{r, r\to \neg a\}, \lozenge_y\rangle$} edge from parent[<-]
					child { node {$\langle\{y, (y\to \neg r) \}, \lozenge_w\rangle$} edge from parent[<-]
						child { node {$\langle\{r, (r\to\neg y)\}, \lozenge_y\rangle$} edge from parent[<-]}}}
				child { node {$\langle\{r, (r\to h)\to (\neg h\to\neg s)\to (\neg s\to\neg y)\}, \lozenge_y)\rangle$} edge from parent[<-]};
				\end{scope}
				\end{tikzpicture}}
			\caption{Argumentation Tree for $\mathcal{VAF}_{KB_1}$}
			\label{fig:arg_tree_2}
		\end{figure}

		The changes $KB_1$ introduces are reflected in a new argumentation tree; that of Figure \ref{fig:arg_tree_2}. A lighter notation has been adopted in Figure \ref{fig:arg_tree_2}, by making $\lozenge_y:=\neg(a\land(a\to y)$ and $\lozenge_w:=\neg(r\land(r\to\neg a)$. For the sake of brevity, instead of deriving every element in $\mathcal{VAF}_{KB_1}$, let us simply note that, while the sets of values, and audiences from $\mathcal{VAF}_{KB_1}$ are preserved, there are two new counter-arguments to the root, altering the value mappings between $V$ and $P$. One affirming that the dedication of land to agricultural activities negatively impacts hydrological regulating services, reducing the water supply, the agricultural yield, and eventually eroding the livelihoods of the locals, i.e., 
		
		$$A_5:=\langle\{r, (r\to h)\to (\neg h\to\neg s)\to (\neg s\to\neg y)\}, \lozenge_y\rangle,$$
		
		and another stating that by impacting the water supply through regulating services, agricultural activities could eventually reduce the agricultural yields, i.e.,
		
		$$A_6:=\langle\{a, a\to \neg h, (\neg h\to\neg s)\to (\neg s\to\neg y)\}, \lozenge_y\rangle,$$
		
		an argument which will not be factored into the derivation of the preferred extension of $\mathcal{VAF}_{KB_1}$ ---hence, the dashed contours-- but one that will serve to motivate our discussion about the non-classical counterpart of the approach embodied in this example.
		
		\begin{figure}[h]
			\centering
			\begin{tikzpicture}[sibling distance=6cm, line width=.3mm]]
			\node[circle, draw, label={\small $a\to y$}] (o) at (0,0) {$a$};
			\node[circle, draw, label={\small $(h\to s)\to(s\to y)$}] (p) at (2.5,0) {$r_y$};
			\node[circle, draw, label={below:\small $h\to w$}] (q) at (1.25,-2) {$r_w$};
			\node[fill=white,inner sep=.5pt, line width=2pt] at (1,1.5) {audience-$y$:};
			\draw[-] (o) to (q);
			\draw[-] (p) to (q);
			\end{tikzpicture}
			\hspace{2cm}
			\begin{tikzpicture}[sibling distance=6cm, line width=.3mm]]
			\node[circle, draw, label={\small $h\to w$}] (r) at (0,0) {$r_w$};
			\node[circle, draw, label={\small $(h\to s)\to(s\to y)$}] (s) at (2.5,0) {$r_y$};
			\node[circle, draw, label={below:\small $a\to y$}] (t) at (1.25,-2) {$a$};
			\node[fill=white,inner sep=.5pt, line width=2pt] at (1,1.5) {audience-$w$:};
			\draw[-] (r) to (t);
			\draw[-] (s) to (t);
			\end{tikzpicture}
			\caption{Hasse Diagram for $\mathcal{VAF}_{KB_1}$}
			\label{fig:hasse_2}
		\end{figure}
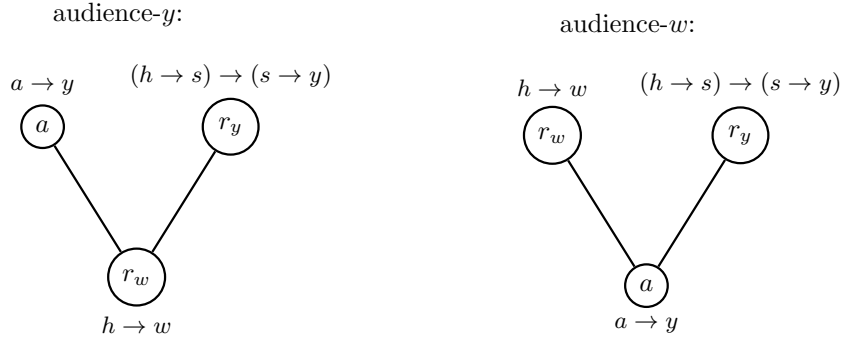{}
		
		Like before, Hasse diagrams are used to represent preferred extensions. The Hasse diagrams of $\mathcal{VAF}_{KB_1}$ appear in Figure \ref{fig:hasse_2}. For audience-$y$, the preferred extension is $\{A_1, A_5\}$. For audience-$w$, the preferred extension is $\{A_2, A_5\}$. The observation that restoration has a beneficial but indirect effect on agricultural yields, reconciles the two audiences through argument $A_5$. The final result being that restoration activities should be prioritised over agricultural ones.
		
		This example comes to an end with a passing remark on argument $A_6$. Argument $A_6$ affirms that an insistence on promoting agricultural activities may be self-defeating, in view of the interdependency of the circumscribing ecological processes. Put another way, $A_6$ introduces the non-trivial thesis $IC:=\{a\land\neg a\}$. 
		
		Given Definition \ref{def:conflict_aud}, it would not be possible to find an admissible argument for either audience-$y$ or audience-$w$ if theses like $IC$ were allowed into $\mathcal{VAF}_{KB_1}$. The reason behind this is that theses in the spirit of $IC$ involve determining the truth value of contradictions, and the acceptance of paradoxes as proofs, tasks for which the numerous classical semantics are not equipped \cite{buning1999propositional}. Since inconsistency arises even in the commonplace observation that ecosystem services are interlinked, recasting VAFs to allow for non-trivial inconsistencies intends to explicate logically a non-negligible part of the complexity of valuating ecosystem services.
	\end{example}
	
	Example \ref{ex:main} shows the advantages of logically modelling the formation of arguments and their critical probing in the context of Argumentative Valuation. The example, however, does not present a complete application of Argumentative Valuation, in that it does not reveal a definite set of values. The reason for this is twofold, the proposed argumentation approach is preliminary, and the trade-off values are already enmeshed in the production of its corresponding VAF's preferred extensions.
	
	The tentative quality of the conceptual framework here presented, is not a matter of incremental development, or a case of missing analytical pieces in the making. Argumentative Valuation is incomplete, for its non-classical logical rooting \textit{is} its explanatory power. Even though VAFs furnish the environment where high-level representations of values and preferences take hold, the logical and concrete antinomies of conflicting forms of fruition of the Páramo can only be fully appreciated ---I conjecture--- through a logical account of their inconsistency ---or rather their para-consistency. Thus, Example \ref{ex:main} is an effort to thematise this conjecture, and by doing so it also showcases some of the constitutive elements of Argumentative Valuation, for the set of defensible arguments it entails, and their connected actions, operate on and prefigure their underlying trade-off values.
	
	\section{Towards (Dialetheic) Argumentative Valuation}
	\label{sec:par_logics}
	
	\say{No proposition is both true and false} \cite{reid2000inquiry}. This is Reid's interpretation of the Principle of Non-Contradiction (PnC), whose quality of being the \textit{firmissimum omnium principiorum} ---Berto reminds us \cite{berto2007sell}--- has made it the supreme cornerstone of knowledge and science. An indubitably common-sensical statement on the actuality of knowledge which seems to require no defence. 
	
	However, contradictions abound in our experiencing of reality, breaking PnC but, ostensibly, not reality. This sense of objectivity in the appearance of contradictions is frequent in situations where interactions are heavily mediated, such as the valuation of ecosystem services, as illustrated in Example \ref{ex:main}. That is, theses involved in the reckoning of ecosystem values ---which lest not forget will be circumscribed to a VAF--- commonly take the form $\{a \land \neg a\}$, for some logical constant or proposition $a$ and the connective conjunction $\land$, e.g.,
	\begin{quote}
		\textit{increased agricultural land-use is desirable and dependent on increments in the water supply which, in turn, depend on hydrological regulating services requiring changes in the use of land away from agricultural practices;}
	\end{quote}
	or equivalently, \textbf{increased agricultural land-use and its negation occur within the same knowledge base}. 
	
	\subsection{Dialectical Set Theory}
	
	The problem with having both $a$ and $\neg a$ as our premises, in a classical logical system, is that any other unrelated premise $b$ can be inferred from them, i.e., $\{a \land \neg a\}\vdash b$, meaning that from inconsistent premises anything follows thus provoking the explosion of the system's information. More situatedly, it could be affirmed that the premises \textit{increased agricultural yield} and \textit{no agricultural activity} lead to the conclusion that \textit{the Páramo can host sportive activities}. That is, the situation concerning sportive activities has been correctly derived from our initial premises, despite being irrelevant to the situation concerning agricultural practices. 
	
	Logical frameworks where this explosion of information is not valid, i.e. where contradictions may appear without the framework's trivialisation, are called paraconsistent. To implement AV we not only need a paraconsistent logic that does not endorse the principle of explosion, we need one where contradictions can be true and employed as building blocks for theorem-proving. In particular, we need a paraconsistent logic that allows for \textit{dialetheias}, i.e., premises such that both themselves and their negations are true \cite{priest2007paraconsistency}, e.g. where it holds that \textit{increased agricultural yield} and \textit{reduced agricultural practices} are valid. 
	
	A simple dialectical set theory (DST) to deal with dialetheias is given by a first-order non-classical theory with a conventional set of connectives $\{\land, \lor, \neg, \to\}$ and universal and particular quantifiers $U$ and $P$, with denumerable stocks of subject variables and some predicate constants \cite{weber2013notes}. The formation rules of DST will be like those for other first-order set theories. The postulates of DST will be those of its non-classical quantificational logic together with some characteristic set-theoretic postulates constraining the predicate constants. 
	
	The quantificational structure of DST (DKQ) is defined by the following axiom scheme for predicate variables $A,B,C$ and $D$ \cite{weber2013notes}:
	\begin{itemize}[label={}]
		\itemequation{}{ A\to A \label{eq:identity}}
		\itemequation{}{ A\land B\to A}
		\itemequation{}{ A\land B\to B}
		\itemequation{}{ A\land (B\lor C)\to (A\land B)\lor C}
		\itemequation{}{ A\land (B\lor C)\to (A\land B)\lor C}
		\itemequation{}{ (A\to B)\land (B\to C)\to (A\to B\land C)}
		\itemequation{}{ (A\to B)\land (B\to C)\to (A\to C)}
		\itemequation{}{ (A\to \neg B)\to (B\to \neg A)	}
		\itemequation{}{ \neg\neg A\to A	}
		\itemequation{}{ A\lor\neg A}
		\itemequation{}{ A\to A(t/x)}
		\itemequation{}{ (A\to B) \to (A\to (x)B)}
		\itemequation{}{ (A\lor B)\implies (A\lor (x)B)}
	\end{itemize}
	alongside the rules listed below:
	\begin{itemize}[label={}]
		\itemequation{}{ A, A\to B\implies B}
		\itemequation{}{ A, B\implies A\land B}	
		\itemequation{}{ (A\to B)\to (C\to D)\implies (B\to C)\to (A\to D)}
		\itemequation{}{ A\implies (x)A \label{eq:last_rule}}
	\end{itemize}
	
	Equations \eqref{eq:identity}--\eqref{eq:last_rule} will determine the validity of a counter-argument within our argumentation framework. That is, an attack or defense argument will be drawn based on DST's axiom scheme and rules. Once drawn, the extension of choice can be applied to the sets of arguments making up the VAF's attack relation. Such is the core idea behind our dialogical approach to paraconsistent logics. 
	
	\subsection{Dialogical Paraconsistency}
	
	There exist at least two popular representations of logic-based argumentation. One that follows the premises and overall structure of Argumentation Theory, and a less intuitionistic alternative rooted in logics. The former corresponds to the tree-like structures of abstract argumentation frameworks (TA) \cite{amgoud2010formal} and the other to Rahman and Carnielli's Literal Dialogues (LD)  \cite{rahman2000dialogical}. We will use both representations interchangably to infuse our argumentative approach to valuation with DST.
	
	LD gives a complete account of dialogues in the form of a series of interactions between two agents ---the Proponent and the Opponent. The Proponent puts forward an argument which will try to defend against all possible attacks of the Opponent. Hence, the argument will be valid iff the Proponent succeeds in its defence, given the global and local rules governing their interactions.

	\begin{table}[h]
		\centering
		\begin{tabular}{m{0.2\textwidth}|m{0.7\textwidth}}
			\hline
			\textbf{Rule}  &   \textbf{Definition} \\\hline
			Dialogue Starter	&  The Proponent begins by asserting a thesis.\\\hline
			Move Order & Agents alternate their moves.\\\hline
			Classical Attack  &  Each agent may attack a statement asserted by its counterpart or defend its own against the latter agent's attack, including the last not already defended assertion.\\\hline
			Intuitionistic Attack  & Each agent may attack a statement asserted by its counterpart or defend its own against the last not already defended attack only.\\\hline
			Negative Literal Attack  & The Proponent is allowed to attack the negation of an atomic (propositional) statement iff the Opponent has already attacked the same statement before.\\ \hline
			Winning Criterion  & If an agent cannot make generate new assertions ---without producing repetitive moves--- the other agent is said to have won the dialogue.\\
			\hline
		\end{tabular}
		\caption{LD's Global Rules \cite{rahman2000dialogical}}
		\label{tab:ld_grules}
	\end{table}
	
	The global rules consist of the principles upon which LD dialogues are structured, as presented in Table \ref{tab:ld_grules}. Local rules are summarised in Table \ref{tab:ld_lrules}. They describe the symbolism for signaling the actions that Opponents and Proponents may take in their role as attackers or defenders with respect to the connectives in DST. These rules grant considerable fluidity to LD. Example \ref{ex:para_ld}, illustrates LD's rules at work under Lorenz's $D\langle 1,1 \rangle$ \cite{rahman2000dialogical} interpretation of LD ---a paraconsistent, although not dialectic, dialogue.
	
	\begin{table}[h]
		\centering
		\begin{tabular}{c|c|c}
			\hline
			$\land, \lor, \neg, \to$  &   \textbf{Attack}  &   \textbf{Defence} \\\hline
			$\neg  A$   &   $A$   &   $\otimes$ \\\hline
			$A\land B$  &  $?L/?R$ (attacker's choice)  &  $A/B$ \\\hline
			$A\lor B$  &  $?$  &  $A/B$ (defender's choice) \\\hline
			$A\to B$  &  $A$  &  $B$ \\
			\hline
		\end{tabular}
		\caption{LD's Local Rules \cite{rahman2000dialogical}}
		\label{tab:ld_lrules}
	\end{table}

	\begin{example}[\textbf{A Paraconsistent Dialogue in LD} \cite{rahman2000dialogical}]
		\label{ex:para_ld}
		Let the Proponent put forward the inconsistent thesis $\{a,\neg a\}$ as in Table \ref{tab:cons_example}, i.e., action $(0)$. The Opponent responds by consecutively inquiring about the right and left sides of the thesis, i.e., actions $?R$ and $?L$ labelled as $(1)$ and $(2)$. Then, the Opponent attacks with premise $a$, i.e., action $(5)$ reacting to action $(2)$, which makes the thesis refutable as the Opponent is the only one of the two who may use an atomic statement ---as per the original/classical set of global rules. If only a single repetition of atomic statements is allowed per agent ---namely $D\langle 1,1 \rangle$---, a form of paraconsistency is introduced as shown in Table \ref{tab:incons_example}. In this case the progression of actions is the same, however after the Proponent has responded to the Opponent's queries the latter is left with no atomic statement to refute the Opponent's.
		
		\begin{table}[h]
			\centering
			\begin{tabular}{ ccc | ccc }
				\multicolumn{3}{c}{Opponent} & \multicolumn{3}{c}{Proponent} \\\hline
				&       &       &       &   $a\land\neg a$   &   (0)\\
				(1)   &   ?R  &   0   &       &       $\neg a$     &   (2)\\
				(3)   &   ?L  &   1   &       &       $a$          &   (4)\\
				(5)   &   $a$ &   2   &       &       $\otimes$    &\\ \hline
				\multicolumn{6}{c}{The Opponent wins}
			\end{tabular}
			\caption{Arguing an Inconsistent Thesis in LD with Classical Rules}
			\label{tab:cons_example}
		\end{table}
		
		\begin{table}[h]
			\centering
			\begin{tabular}{ ccc | ccc }
				\multicolumn{3}{c}{Opponent} & \multicolumn{3}{c}{Proponent} \\\hline
				&       &       &       &   $a\land\neg a$   &   (0)\\
				(1)   &   ?R  &   0   &       &       $\neg a$     &   (2)\\
				(3)   &   ?L  &   1   &       &       $a$          &   (4)\\\hline
				\multicolumn{6}{c}{The Proponent wins}
			\end{tabular}
			\caption{Arguing an Inconsistent Thesis in LD under $D\langle 1,1\rangle$}
			\label{tab:incons_example}
		\end{table}
	\end{example}
	
	$D\langle 1, 1\rangle$ provides a very limited view of paraconsistency. In particular, simple formulas such as $a\to (\neg a\to b)$ retain their validity in $D\langle 1, 1\rangle$, trivialising our logical model. To cater for this more indirect forms of trivialisation, the negative literal rule has been introduced, thereby recognising that there is some contexts in which $a$ and $\neg a$ can be asserted. To see this let us further formalise our dialogical logic framework, so we can make use of DST.
	
	\subsection{Dialogical Logic} 
	
	Dialogical Logic is based on a first-order dialogical language where every term is either a variable or an individual constant. Dialogical Logic's formalisation builds on the work of \cite{Clerbout2014}, and its notation is borrowed from \cite{beirlaen2016inconsistency}. However, the interplay between structural and local rules follows Table \ref{tab:ld_grules} and Table \ref{tab:ld_lrules}. Dialogical Logic is formally stated in Definition \ref{def:dialog_log}.
	
	\begin{definition}[\textbf{Dialogical Logic} \label{def:dialog_log}\cite{Clerbout2014}]
		Let $\mathcal{L}$ be a propositional language such that $\phi ::= \phi |\phi\land\phi |\phi\lor\phi |\phi\to\phi | \neg\phi$. The first lower-case letters in the alphabet $a,b,\ldots$ designate logical constants and lower-case letters $p,q,r,\ldots$ refer to atomic formulas in $\mathcal{L}$. Lower-case greek letters $\phi, \psi, \chi, \ldots$ denote $\mathcal{L}$-formulas, while upper-case Greek letters $\Gamma, \Sigma, \Delta, \ldots$ designate finite sets of $\mathcal{L}$-formulas. Arguments take the form of atomic and $\mathcal{L}$-formulas. 
	\end{definition}
	
	The notion of dialogue requires more structure specifying the interactions between the Proponent and the Opponent.  \textbf{P}  and \textbf{O} are used to label the arguments put forward by the Proponent or the Opponent, who may also be designated through the placeholders \textbf{X} and \textbf{Y} with \textbf{X}$\neq$\textbf{Y}, whenever their identity becomes secondary. A formal characterisation of dialogues is given in Definition \ref{def:dialog}. 
	
	\begin{definition}[\textbf{Dialogue}\label{def:dialog}]
		A dialogue having $\psi[\phi_0,\ldots,\phi_{n-1}]$ as its initial thesis, i.e., beginning with the claim that the conclusion $\psi$ follows from the premises $\phi_0,\ldots,\phi_{n-1}$, is the set $\mathcal{D}(\psi[\phi_0,\ldots,\phi_{n-1}])$ of moves performed by \textbf{O} and \textbf{P}. 
		
		Moves are assertions or requests which may serve to attack ($A$), defend ($D$) or inquire about a particular premise. Moves are represented as expressions of the form \textbf{X-}$e$, where \textbf{X} is as before and $e$ stands for either an assertion or a request. The symbols \say{!} and \say{?} signal the agents' assertions and requests, respectively. 
		
		Agents' ranks $r_i\in\mathbb{N}, i\in\{1,2\}$ with \textbf{O}$\to 1$ and \textbf{P}$\to 2$ indicate the number of attacks and defences they can play within the dialogue $\mathcal{D}(\psi[\phi_0,\ldots,\phi_{n-1}])$. Agents assert their ranks as so: \textbf{O-}$n:=r_1$ and \textbf{P-}$m:=r_2$. The counters of moves in $\mathcal{D}$ are denoted by $P_{\mathcal{D}}(\cdot)$, and are also referred to as the dialogue's position.
	\end{definition}
	
	In this formalism, the rules in Table \ref{tab:ld_grules} and Table \ref{tab:ld_lrules} can be rewritten as in Table \ref{tab:dialog_grules} and Table \ref{tab:dialog_lrules}, respectively. Their interpretation is preserved as well as their relative inadequacy for capturing dialetheis 
	
	\begin{table}[h]
		\centering
		\begin{tabular}{P{0.12\textwidth}|P{0.8\textwidth}}
			\hline
			\textbf{Rule}  &   \textbf{Definition} \\\hline
			Dialogue Starter	&  \specialcell{$P_{\mathcal{D}}(\textbf{P-}!\psi[\phi_0,\ldots,\phi_{n-1}])=0$,\\ $P_{\mathcal{D}}(\textbf{P-}n:=r_1)=1$ and $P_{\mathcal{D}}(\textbf{P-}n:=r_2)=2$}\\\hline
			Move Order  &  \specialcell{$F_{\mathcal{P}}(M)=[m, Z], m<P_{\mathcal{D}}(M), Z\in\{A,D\}$\\ \specialcell{$[\mathcal{P}]_{-1} = \textbf{Y-}e, M_0\equiv[\mathcal{P}]_{0} = \textbf{Y-}e$},\\ there are $n$ $\textbf{X-}!e$ moves s. t. $F_{\mathcal{P}}(M_0)=\ldots=F_{\mathcal{P}}(M_{n-1})=[m_0, Z], Z\in\{A,D\}$}, and a move $N = \textbf{X-}f$ s. t. $F_{\mathcal{P}\cup\{N\}}(N)=[m_0, Z]$ iff $n<r$ where \textbf{X-}$l:=r$\\\hline
			Classical Attack  &  \specialcell{$N = \textbf{P-}\psi\mathcal{P}$,\\ then there exists $N = \textbf{O-}!\psi\in\mathcal{P}$ such that $P_{\mathcal{P}}(M)<P_{\mathcal{P}(N)}$}\\\hline		
			Winning Criterion  & Agent \textbf{X} wins $\mathcal{P}$ iff there is no move $Q=\textbf{Y-}g$ s.t. $\mathcal{P}\cup\{N\}\in\mathcal{D}$ whenever $[\mathcal{P}]_{-1}=\textbf{X-}e$ \\
			\hline
		\end{tabular}
		\caption{Global Rules for a Formal Dialogue \cite{Clerbout2014}}
		\label{tab:dialog_grules}
	\end{table}
	
	\begin{table}
		\centering
		\begin{tabular}{ccc}
			Assertion & Attack & Defence\\\hline
			\textbf{X-}$!\phi\land\psi$ & \textbf{Y-}$?\land_{L}$/\textbf{Y-}$?\land_{R}$ & \textbf{X-}$!\phi$ /\textbf{X-}$!\psi$\\
			\textbf{X-}$!\phi\lor\psi$  & \textbf{Y-}$?\lor$ & \textbf{X-}$!\phi$/ \textbf{X-}$!\psi$\\
			\textbf{X-}$!\neg\phi$	    & \textbf{Y-}$!\phi$ & --\\    		
			\textbf{X-}$!\phi\to\psi$ & \textbf{Y-}$!\phi$  & \textbf{X-}$!\psi$\\    		
			\textbf{X-}$!\psi[\phi_0,\ldots,\phi_{n-1}]$ & \textbf{Y-}$!\phi_0,\ldots,$\textbf{Y-}$\phi_{n-1}$ & \textbf{X-}$!\psi$\\    		    		
		\end{tabular}
		\caption{Local Rules for a Formal Dialogue \cite{Clerbout2014}}
		\label{tab:dialog_lrules}
	\end{table}
	
	It remains to see whether robust forms of the Negative Literal Attack enable more apt environments to incorporate the scheme in Equations \ref{eq:identity}--\ref{eq:last_rule} into a formal paraconsistent dialogue. The viability of doing so will determine the possibility of generating a dialogical \textit{notion of consequence} ---i.e., a criterion of entailment--- capable of dealing with dialetheias. This, in turn, will condition our capacity to produce semantic tableaux to be used as the building blocks of AV's argumentation solver.
	
	\section{Conclusions and Future Work}
	\label{sec:future}
	
	This proposal argues that computational argumentation and non-classical logics offer an adequate conceptual framework to approach the valuation of ecosystem services in the context of Páramo ecosystems. I refer to this approach as \textit{Argumentative Valuation}. Argumentative Valuation is a participatory valuation framework that uses stated preferences to reason about monetary and non-monetary trade-off values.
	
	The main points made here can be summarised as follows:
	\begin{itemize}[label={},itemsep=1pt,topsep=1pt]
		\item \textbf{Computational argumentation provides an adequate framework to conduct the valuation of ecosystem services, in terms of the critical questioning of the arguments concerning the fruition of said services}. Value-based Argumentation Frameworks incorporate the preferences and values that support a given claim, creating a relational context where trade-offs are made explicit through the semantics delineated by the maximal set of admissible arguments. The resulting preferred extensions, or sets of defensible arguments with respect to this semantical criterion, are manifestations of trade-off values and indicate a particular ordering of the connected actions on the Páramo. 
		\item \textbf{Argumentative Valuation should be founded upon a non-classical model of logical entailment}. The interaction of the qualitative and quantitative aspects of the arguments typically encountered in the valuation of ecosystem services, are better represented as logical constructions. Given that conflicts and inconsistencies ---i.e., non-trivial theses of the form $\{a \land \neg a\}$, where $a$ is a premise belonging to the support of a generic argument $A\in\mathcal{A}$--- configure the mode of relation among agents of valuation, the logical circuitry of Value-based Argumentation Frameworks should be non-classical.
		\item \textbf{Argumentative Valuation allows researchers to become active agents of valuation, and the non-classical approach to their implementation do not require the agents of valuation to be economically rational}. Although these claims were not sufficiently explored, they naturally emanate from the possibilities of a logical framework that does not abide by consistency, and from the fact that the researchers' axiological principles can be directly modelled within the same argumentation framework. 
	\end{itemize}

	\newpage
	Some of the themes left for future work include the following:
	\begin{itemize}[itemsep=1pt,topsep=1pt]
		\item Discerning the theory that upholds my insistence on non-classical logics, and circumscribing its semantics to the theory of Value-based Argumentation Frameworks. 
		\item Incorporating Dialectical Set Theory into Dialogical Logic, to semantic tableaux supporting AV's value-based argumentation framework. 
		\item Outlining the order-theoretical problem inherent to the derivation of the new notions of extension induced by dialetheic semantic tableaux. Given the relative simplicity of Example \ref{ex:main}, the combinatorial considerations involved in the implementation of preferred semantics might have gone unnoticed.
		\item Discussing the nature of value in the context of Argumentative Valuation.
	\end{itemize}{}
	
	\bibliographystyle{alpha}
	\bibliography{arg_val}
	
\end{document}